\documentclass[epj]{svjour}

\usepackage{graphicx}
\usepackage{fancyhdr}
\def\makeheadbox{{
 }}
\def\mytitle{My title} 
\def\myauthors{My name}  
\def\mytype{My type of session}
\def\mysession{My session}

\def\mytitle{Squark and gaugino hadroproduction in NMFV-SUSY}
\def\myauthors{Bj\"orn Herrmann}
\def\mytype{Contributed Talk}    
\def\mysession{Flavor Physics}

\begin{document}

\title{Squark and gaugino hadroproduction and decays in non-minimal flavour violating supersymmetry}

\author{Bj\"orn Herrmann 
}                    

\institute{Laboratoire de Physique Subatomique et de Cosmologie\\
	   Universit\'e Joseph Fourier / CNRS-IN2P3 / INPG, 
	   53 Avenue des Martyrs, 38026 Grenoble, France\\
}

\date{September 11, 2007}

\abstract{
We implement non-minimal flavour violation (NMFV) in the MSSM at low scale. In this
framework, we propose benchmark points for the mSUGRA scenario including
non-minimal flavour violation by evaluating a number of experimental low-energy,
electroweak precision and cosmological constraints.
We finally discuss phenomenological aspects of our NMFV scenario and present a numerical
analysis of squark and gaugino production cross sections at the LHC.
} 

\maketitle

\section{Flavour violation in the MSSM}
\label{sec1}

\vspace*{-12.3cm}LPSC 07-97\vspace*{12.3cm}

While the only source of flavour violation in the Standard Model (SM) arises
through the rotation of the quark interaction eigenstates into the basis of
physical mass eigenstates, in Supersymmetry (SUSY) additional flavour violation
is introduced at the weak scale through renormalization group running
\cite{ref11}. In minimal flavour violating (MFV) scenarios, the non-diagonal
entries of the squared squark mass matrices stem from the trilinear Yukawa 
couplings of the fermion and Higgs su\-per\-mul\-tiplets and the resulting 
different renormalizations of quarks and squarks. In constrained MFV (cMFV)
scenarios, these flavour violating entries are neglected at the SUSY
breaking and weak scale.

\smallskip
However, when SUSY is embedded in larger structures such as GUT theories, new
sources of flavour violation arise \cite{ref15}, that have to be included in the
squared squark mass matrices at the weak scale. As the corresponding
flavour violating off-diagonal entries $\Delta_{ij}^{qq'}$ cannot be simply
deduced from the CKM-matrix alone, in NMFV is parametrized by considering
them as free parameters. Denoting $M_{L_k}^2$,
$M_{R_k}^2$ and $m_kX_k$ their usual diagonal and helicity mixing entries,
$k=1,2,3$ referring to the (s)quark family, the squared squark mass matrices are
then given by 
\renewcommand{\arraystretch}{1.2}
\begin{equation}
    \hspace*{-2.1mm} M_{\tilde{q}}^2\!=\!\!\left( \!\! \begin{array}{ccc|ccc}
 	    M^2_{L_1} & \Delta^{12}_{LL} & \Delta^{13}_{LL} & m_1 X_1 & \Delta^{12}_{LR} & \Delta^{13}_{LR} \\ 
   	    \Delta^{21}_{LL} & M^2_{L_2} & \Delta^{23}_{LL} & \Delta^{21}_{LR} & m_2 X_2 & \Delta^{23}_{LR} \\ 
     	    \Delta^{31}_{LL} &  \Delta^{32}_{LL} & M^2_{L_3} & \Delta^{31}_{LR} & \Delta^{32}_{LR} & m_3 X_3 \\ 
	    \hline 
	    m_1 X_1^* & \Delta^{12}_{RL} & \Delta^{13}_{RL} & M^2_{R_1} & \Delta^{12}_{RR} & \Delta^{13}_{RR} \\
	    \Delta^{21}_{RL}~& m_2 X_2^* &  \Delta^{23}_{RL} & \Delta^{21}_{RR} & M^2_{R_2} & \Delta^{23}_{RR} \\ 
	    \Delta^{31}_{RL}~& \Delta^{32}_{RL} & m_3 X_3^* & \Delta^{31}_{RR} & \Delta^{32}_{RR} & M^2_{R_3}
        \end{array} \!\! \right)\!\!.
\end{equation}

The scaling of the flavour violating entries $\Delta_{ij}^{qq'}$ with the
SUSY-breaking scale $M_{\rm SUSY}$ implies a hierarchy $\Delta_{LL}^{qq'} \gg
\Delta_{RL,LR}^{qq'} \gg \Delta_{RR}^{qq'}$ \cite{ref15}. 
The numerically largest $\Delta_{LL}^{qq'}$ are also related by $SU(2)$ gauge 
invariance through the CKM-matrix, not allowing a large difference between them.
They are usually normalized to the diagonal entries \cite{ref18}, 
\begin{equation}
    \Delta^{qq'}_{ij} = \lambda^{qq'}_{ij} M_{i_q} M_{j_{q'}} ,
\end{equation}
so that NMFV is gouverned by 24 arbitrary dimensionless parameters
$\lambda^{qq'}_{ij}$.

The diagonalization of the mass matrices $M_{\tilde{u}}^2$ and $M_{\tilde{d}}^2$
requires the introduction of two additional $6\times 6$ matrices $R^u$ and
$R^d$, which relate the helicity eigenstates to the physical mass eigenstates
through
\begin{eqnarray}
    (\tilde{u}_1, \dots, \tilde{u}_6)^T &=& 
    R^u (\tilde{u}_L, \tilde{c}_L, \tilde{t}_L, \tilde{u}_R, \tilde{c}_R, \tilde{t}_R)^T , \\
    (\tilde{d}_1, \dots, \tilde{d}_6)^T &=& 
    R^d (\tilde{d}_L, \tilde{s}_L, \tilde{b}_L, \tilde{d}_R, \tilde{s}_R, \tilde{b}_R)^T . 
\end{eqnarray}
By convention, the squark mass eigenstates are labelled according to 
$m_{\tilde{q}_1} < ... < m_{\tilde{q}_6}$ for $q=u,d$. All relevant couplings
are then generalized by expressing them in terms of $R^u$ and $R^d$.

\section{Constraints on NMFV in mSUGRA}
\label{sec2}

In order to take implicitly into account a number of experimental bounds
coming from the neutral kaon sector, $B$- and $D$-meson oscillations, rare
decays, and electric dipole moments \cite{ref34,ref36}, we restrict
ourselves to the case of flavour violation between only the second and third
generations of left-chiral squarks, implemented through one real NMFV-parameter
$\lambda \equiv \lambda_{LL}^{sb} = \lambda_{LL}^{ct}$, while all other
$\lambda_{ij}^{qq'}$ are zero. Allowed regions for this parameter are then
obtained by imposing explicitly low-energy, electroweak precision and
cosmological constraints. We start with the theoretically robust inclusive
branching ratio \cite{ref37}
\begin{equation}
    {\rm BR}(b\to s\gamma) = (3.55 \pm 0.26) \times 10^{-4},
\end{equation}
which affects directly the allowed squark mixing bet\-ween the second and third
generation. Since NMFV contributes already at one-loop level, as do the SM
contributions, this constraint is supposed to depend strongly on the
NMFV-parameter $\lambda$.
A second observable, sensitive to squark mass splitting and thus to NMFV, is the
electroweak $\rho$-parameter, for which new physics contributions are
constrained to \cite{ref33}
\begin{equation}
    \Delta\rho = -\alpha T = (1.02 \pm 0.86) \times 10^{-3} .
\end{equation}
by the latest combined precision measurements. A third variable sensitive to 
SUSY loop corrections is the ano\-malous magnetic moment of the muon, for which
we require the new physics contribution $a_{\mu}^{\rm SUSY}$ to close the gap
\begin{equation}
    \Delta a_{\mu} = (22 \pm 10)\times 10^{-10} 
    \label{eqamu}
\end{equation}
between recent experimental data and the SM prediction \cite{ref33}.
Finally, to have a viable cold dark matter candidate, we require the lightest
SUSY particle (LSP) to be neutral in electric charge and colour, and the
resulting neutralino relic density $\Omega_{\rm CDM}h^2$ to agree with the
observational limit \cite{ref44} 
\begin{equation}
    0.094 < \Omega_{\rm CDM}h^2 < 0.136 .
    \label{eqcdm}
\end{equation}

The above limits are imposed at $2\sigma$ confidence level on the mSUGRA model
with five free parameters $m_0$, $m_{1/2}$, $A_0$, $\tan\beta$, and sgn($\mu$)
at the GUT scale. The renormalization group equations are solved numerically to
two-loop order using the computer programme SPheno 2.2.3 \cite{ref46}. At the
weak scale, we generalize the squark mass matrices by introducing the NMFV
parameter $\lambda$ as described above, diagonalize these matrices and
compute the low-energy and electroweak precision constraints using the computer
programme FeynHiggs 2.5.1 \cite{ref47}. The neutralino relic density is
computed using a NMFV-adapted version of the computer programme DarkSUSY 4.1
\cite{ref43}, taking into account the six-dimensional helicity and flavour
squark mixing. For the Standard Model input parameters, i.e. the masses and
widths of the electroweak gauge bosons and quarks, the elements of the
CKM-matrix, the SM $CP$-violating phase, and Fermi's coupling constant, we
refer the reader to Ref. \cite{ref33}.

The one-loop SUSY contributions to $a_{\mu}^{\rm SUSY}$ are proportional to
sgn($\mu$) \cite{ref48}, so that the disagreement between experiment and SM
prediction is increased for $\mu<0$ in all SUSY models. As furthermore, this
region is virtually excluded by the $b\to s\gamma$ constraint for all
$\lambda\in[0;0.1]$ except for very high SUSY masses, we restrict ourselves to
the case of $\mu>0$. The dependence on $A_0$ of the variables is extremely weak, 
so that we use only $A_0=0$ throughout our analysis. 

\begin{figure*}
    \includegraphics[scale=0.235]{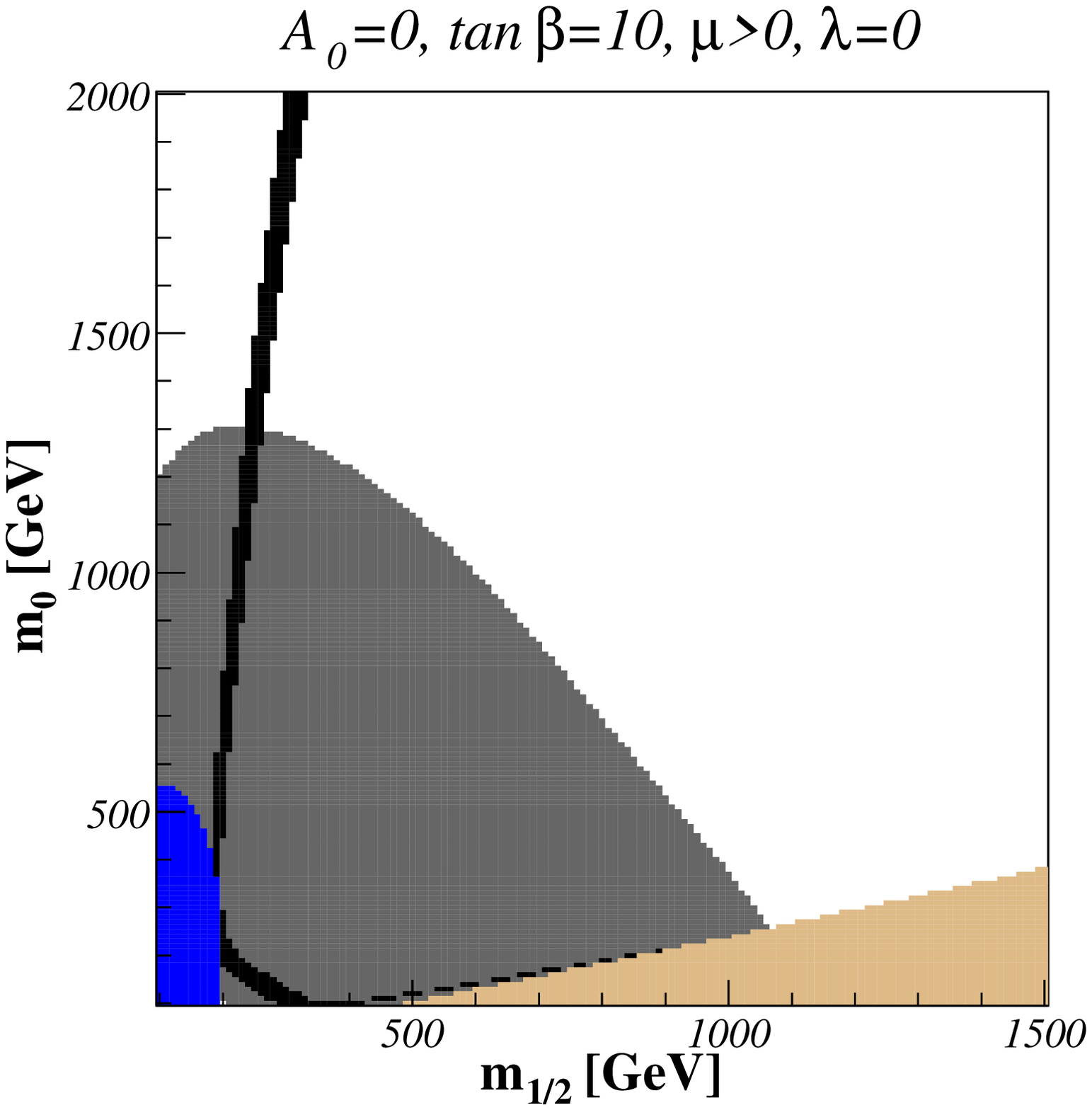}\hspace*{2mm}
    \hfill
    \includegraphics[scale=0.235]{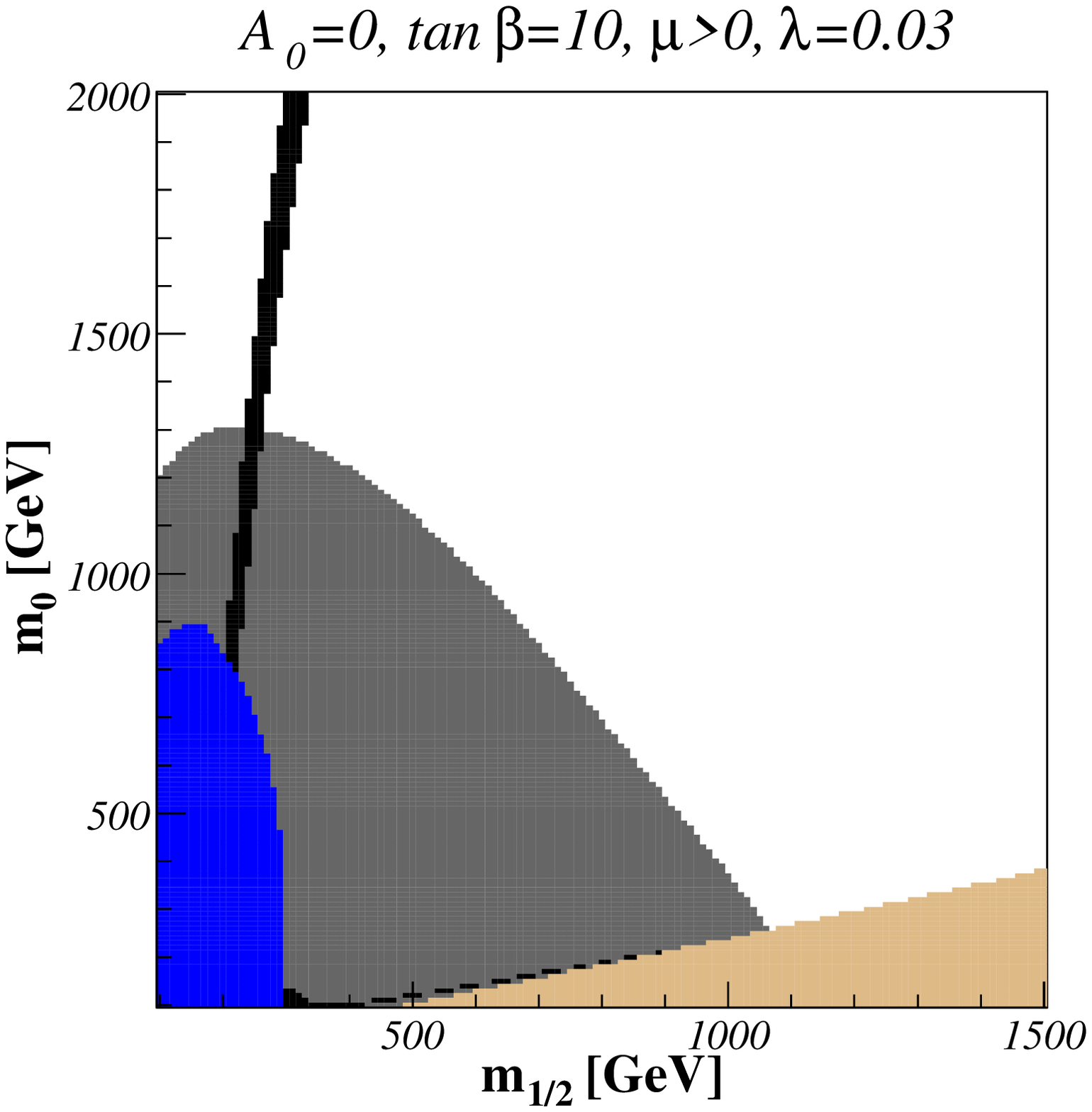}\hspace*{2mm}
    \hfill
    \includegraphics[scale=0.235]{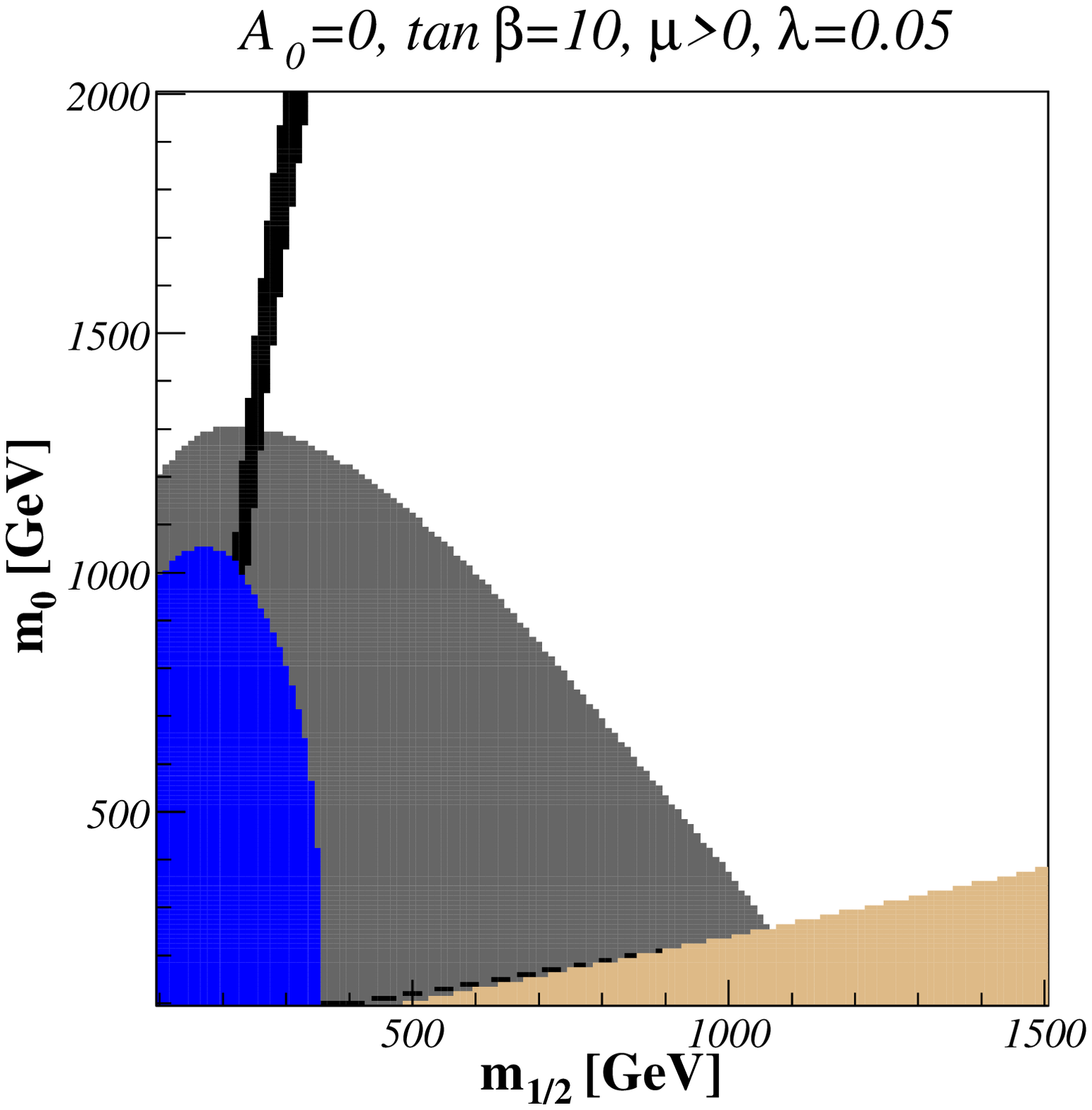}\hspace*{2mm}
    \hfill
    \includegraphics[scale=0.235]{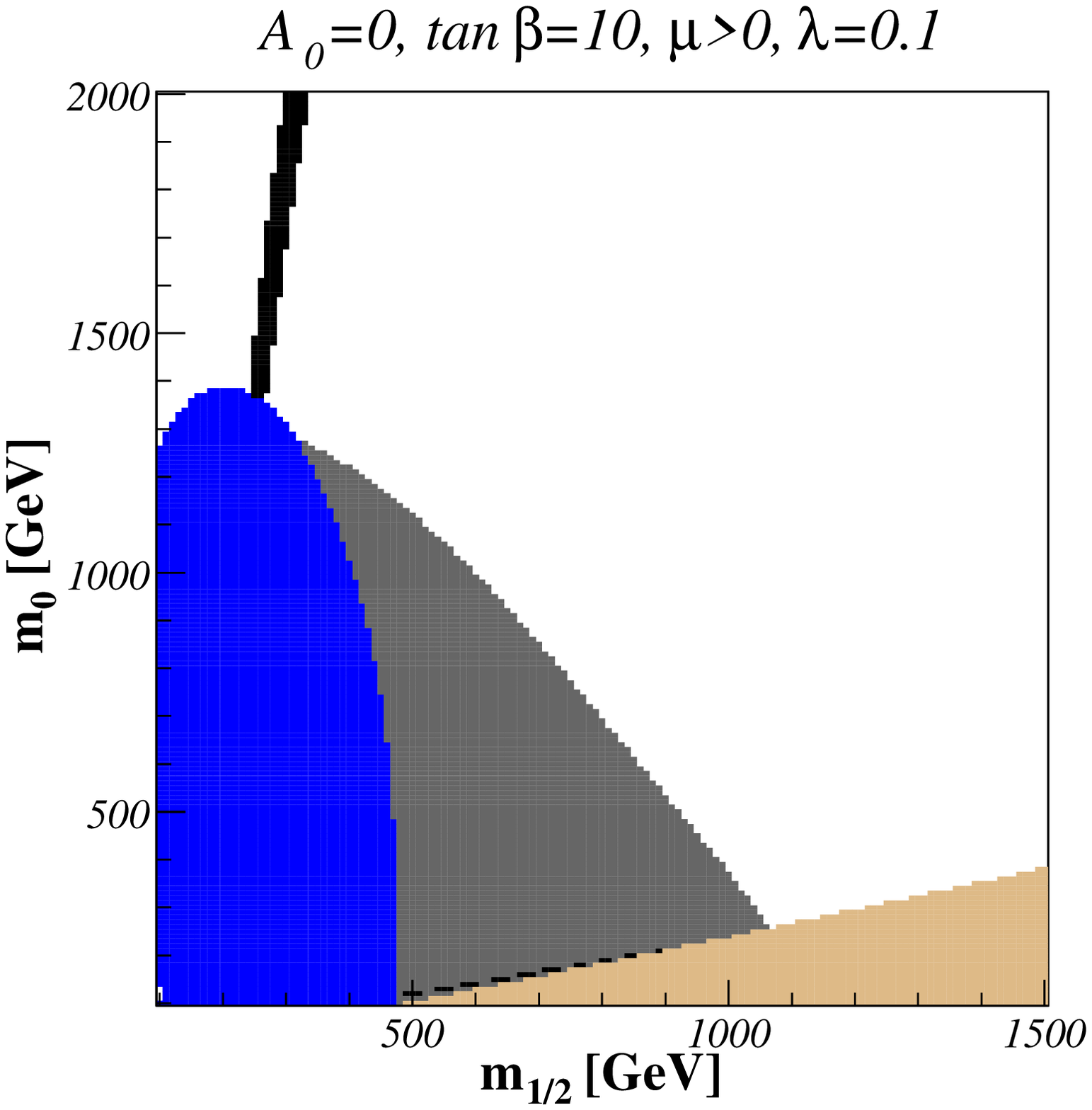}
\caption{The $(m_0,m_{1/2})$-planes for $\tan\beta=10$, $A_0=0$, $\mu>0$, and
$\lambda=0,0.03,0.05,0.1$. We show $a_{\mu}$ (grey) and WMAP (black) favoured as
well as $b\to s\gamma$ (blue) and charged LSP (beige) excluded regions of the
mSUGRA parameter space in constrained minimal ($\lambda=0$) and non-minimal
($\lambda>0$) flavour violation.} 
\label{fig1}
\end{figure*}

Fig. \ref{fig1} shows a typical scan of the mSUGRA parameter space in $m_0$ and
$m_{1/2}$ for $\tan\beta=10$, $A_0$=0, and $\mu>0$, and different values of
$\lambda \in [0; 0.1]$. As expected, the $b\to s\gamma$ 
excluded region depends strongly on flavour mixing, while the regions favoured
by $a_{\mu}$ and the dark matter relic density show quite low sensitivity on the
$\lambda$-parameter. $\Delta\rho$ constrains the parameter space only for very
high scalar ($m_0 > 2000$ GeV) and gaugino ($m_{1/2} > 1500$ GeV) masses, so
that the corresponding regions are not shown.

Based on an extensive analysis of the mSUGRA parameter space at different values
of $\tan\beta$ (10, 30, and 50), we define benchmark scenarios, which are
allowed/favoured by the above constraints, permit non-minimal flavour violation
among left-chiral squarks of the second and third generation up to $\lambda \le
0.1$, and are at the same time ``collider-friendly", i.e. have relatively low
values of $m_0$ and $m_{1/2}$. Our choices are presented in Tab. \ref{tab1},
together with the corresponding allowed ranges for the NMFV-parameter $\lambda$.
Note that all four benchmark points are also valid in MFV ($\lambda \le 0.005
\dots 0.01$) and cMFV ($\lambda=0$) scenarios.
\begin{table}
  \caption{Benchmark points for mSUGRA allowing for flavour violation among the
  second and third generations. We also indicate the allowed range for the
  NMFV-parameter $\lambda$. The values of $m_0$, $m_{1/2}$ and $A_0$ are given
  in GeV.} 
  \label{tab1}
  \begin{tabular}{ccccccc}
  \hline\noalign{\smallskip}
    & $m_0$ & $m_{1/2}$ & $A_0$ & $\tan\beta$ & sgn($\mu$) & $\lambda$ bounds \\
  \noalign{\smallskip}\hline\noalign{\smallskip}
  A & 700 & 200 & 0 & 10 & + & [0;~0.05] \\
  B & 100 & 400 & 0 & 10 & + & [0;~0.10] \\
  C & 230 & 590 & 0 & 30 & + & [0;~0.05] \\
  D & 600 & 700 & 0 & 50 & + & [0;~0.05] \\
  \noalign{\smallskip}\hline
  \end{tabular}
\end{table}

\section{Numerical discussion}
\label{sec3}

In this talk, we focus on the phenomenology of benchmark scenario B. For this point
all sparticle masses are below 1 TeV (see Fig. \ref{fig3}), which makes it the most
``collider-friendly" among the four proposed scenarios, having light
sleptons ($m_{\tilde{l}}\sim 200...300$ GeV), relatively light gauginos
($m_{\tilde{\chi}}\sim 150...550$ GeV) and squarks ($m_{\tilde{q}}\\ \sim 650...850$
GeV), as well as a heavy gluino ($m_{\tilde{\chi}}\sim 900$ GeV). 
For a similar discussion of the other points presented in Tab. \ref{tab1}, the
reader is referred to Ref. \cite{ref0}.

\begin{figure*}
    \includegraphics[scale=0.22]{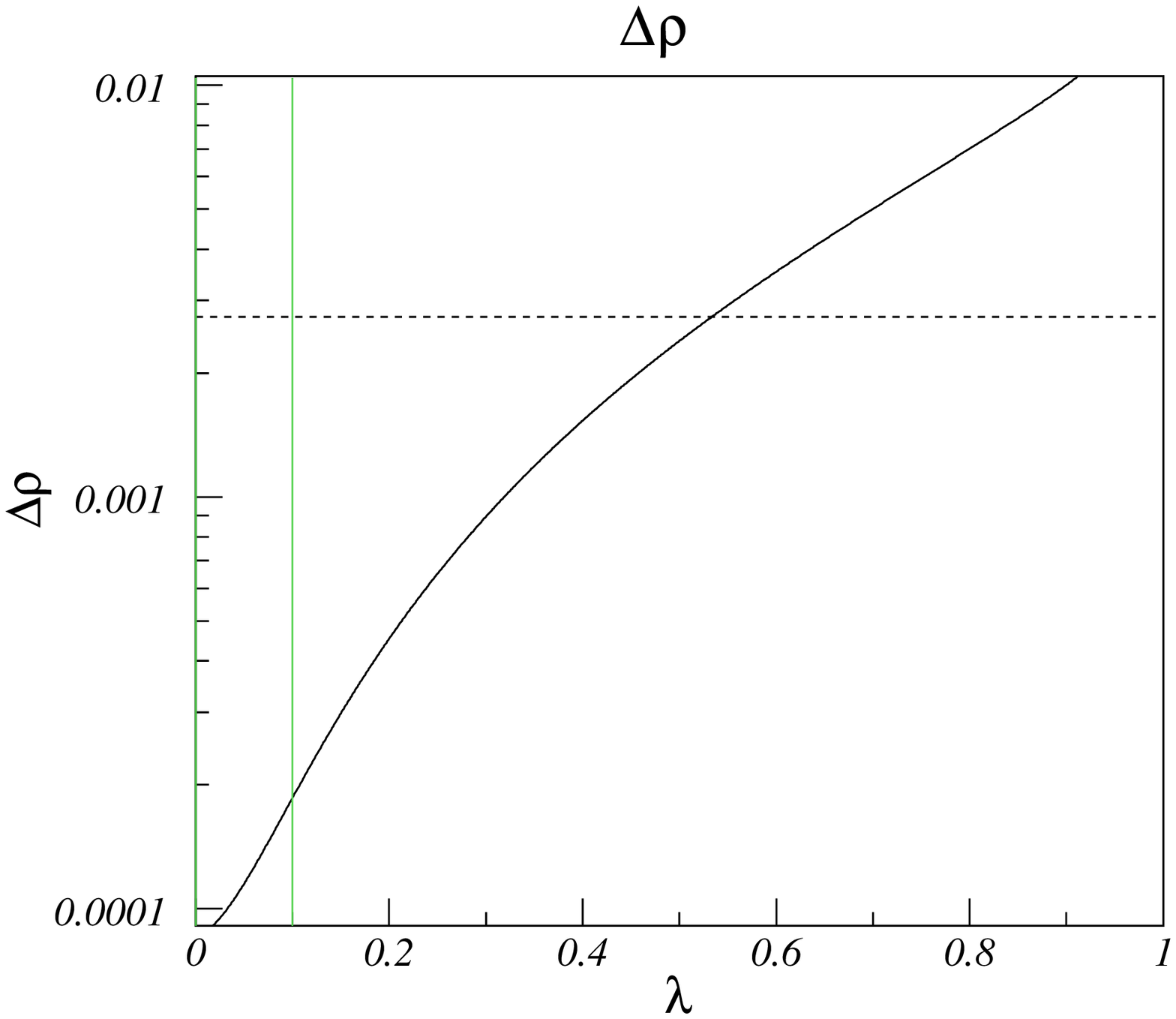}
    \hfill
    \includegraphics[scale=0.22]{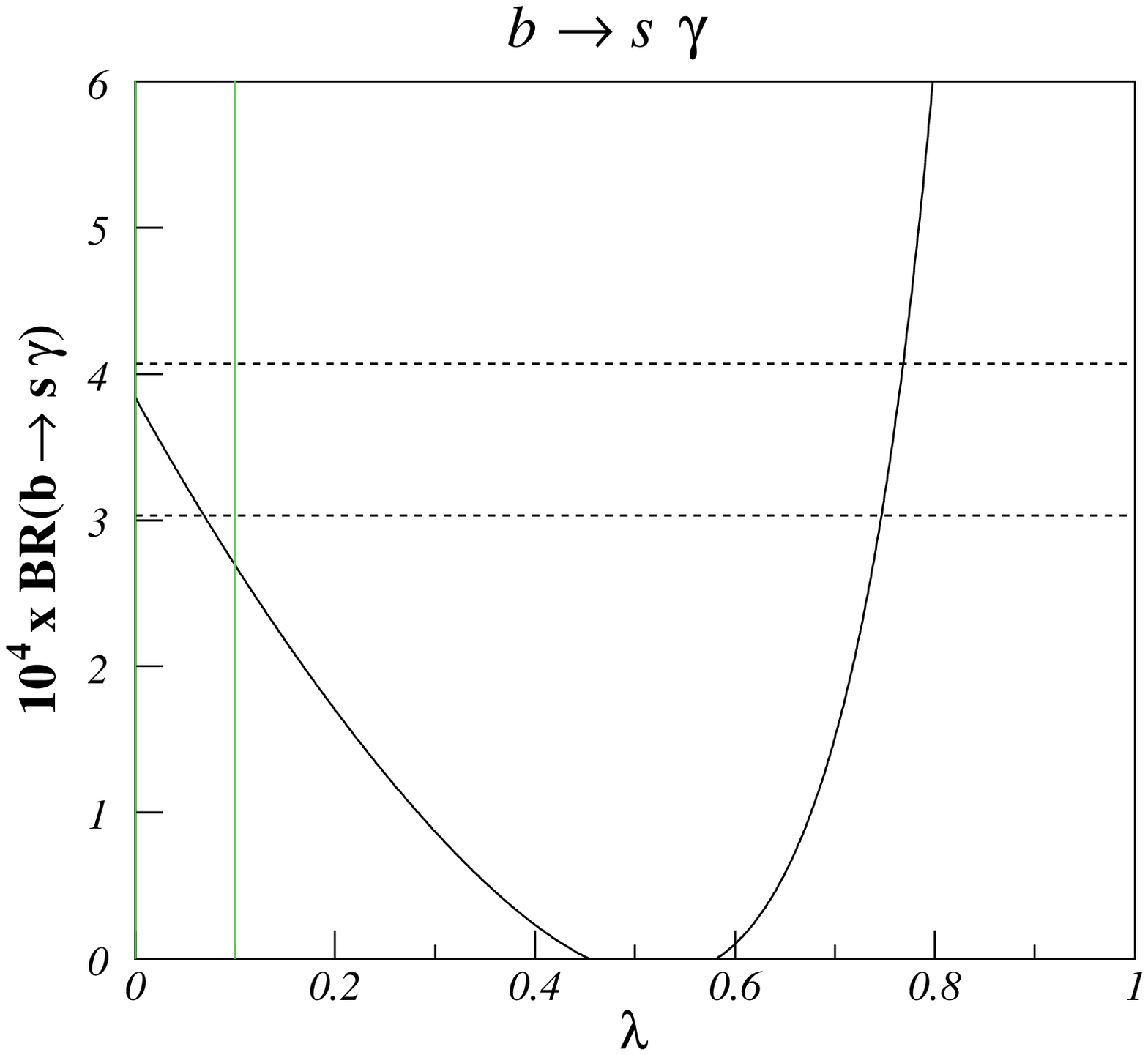}
    \hfill
    \includegraphics[scale=0.22]{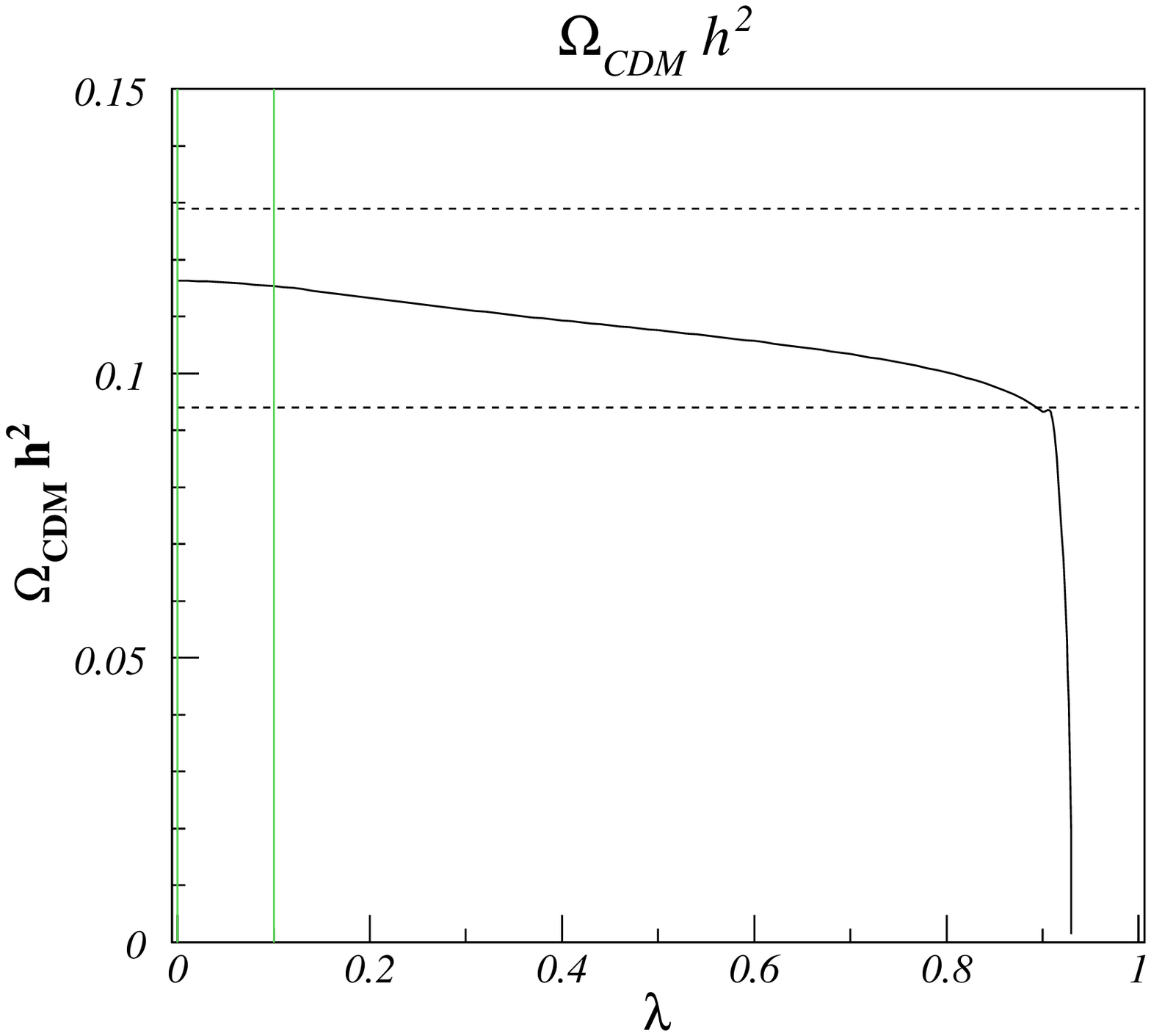}
    \hfill
    \includegraphics[scale=0.22]{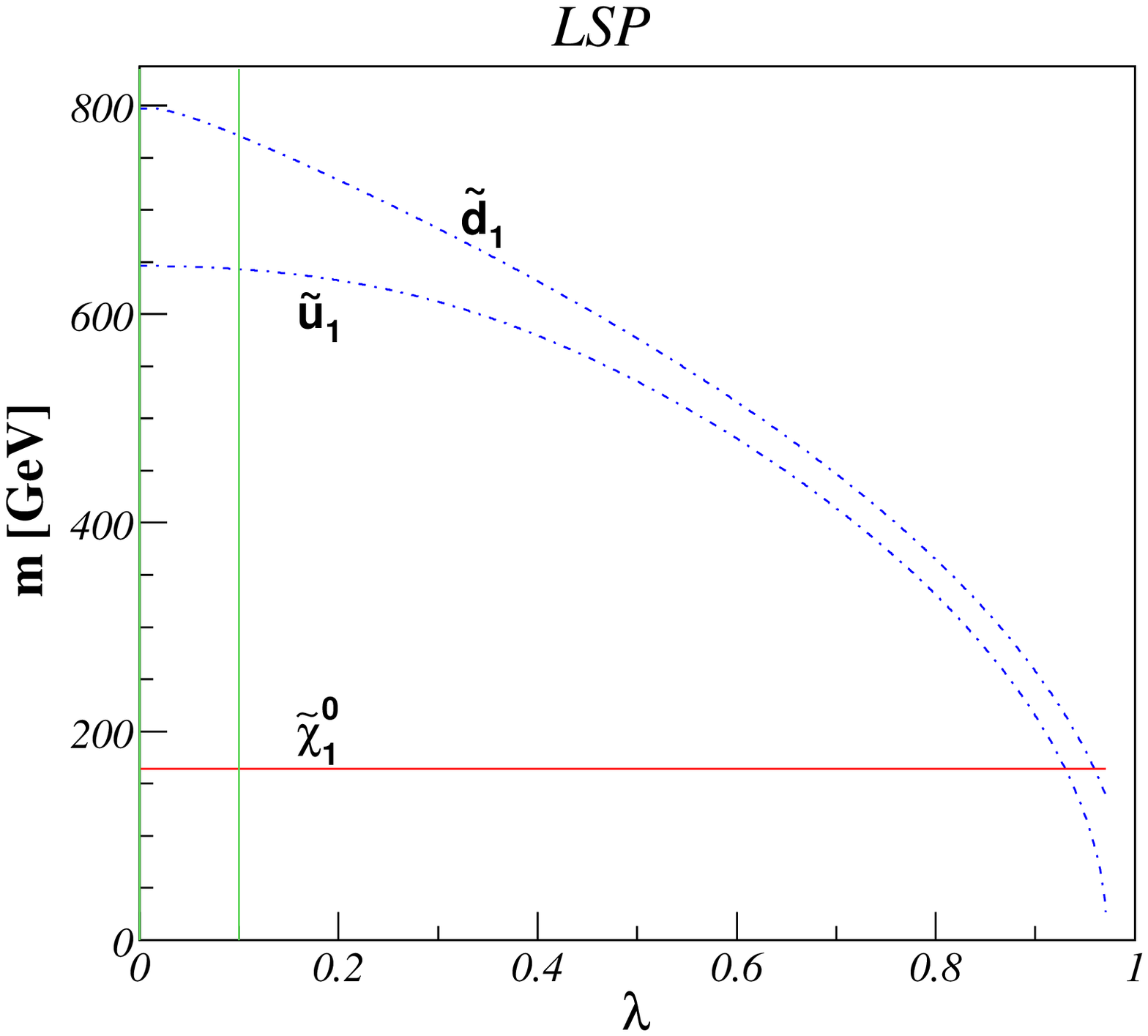}
\caption{Dependence of the precision variables BR($b\to s\gamma$), $\Delta\rho$,
and the cold dark matter relic density $\Omega_{CDM}h^2$ as well as the lightest
neutralino, up- and down-type squark masses on the NMFV-parameter $\lambda$ in
our benchmark scenario B. The experimentally allowed ranges (within 2$\sigma$)
are indicated by horizontal dashed lines.} 
\label{fig2}
\end{figure*}

\begin{figure}
    \includegraphics[scale=0.23]{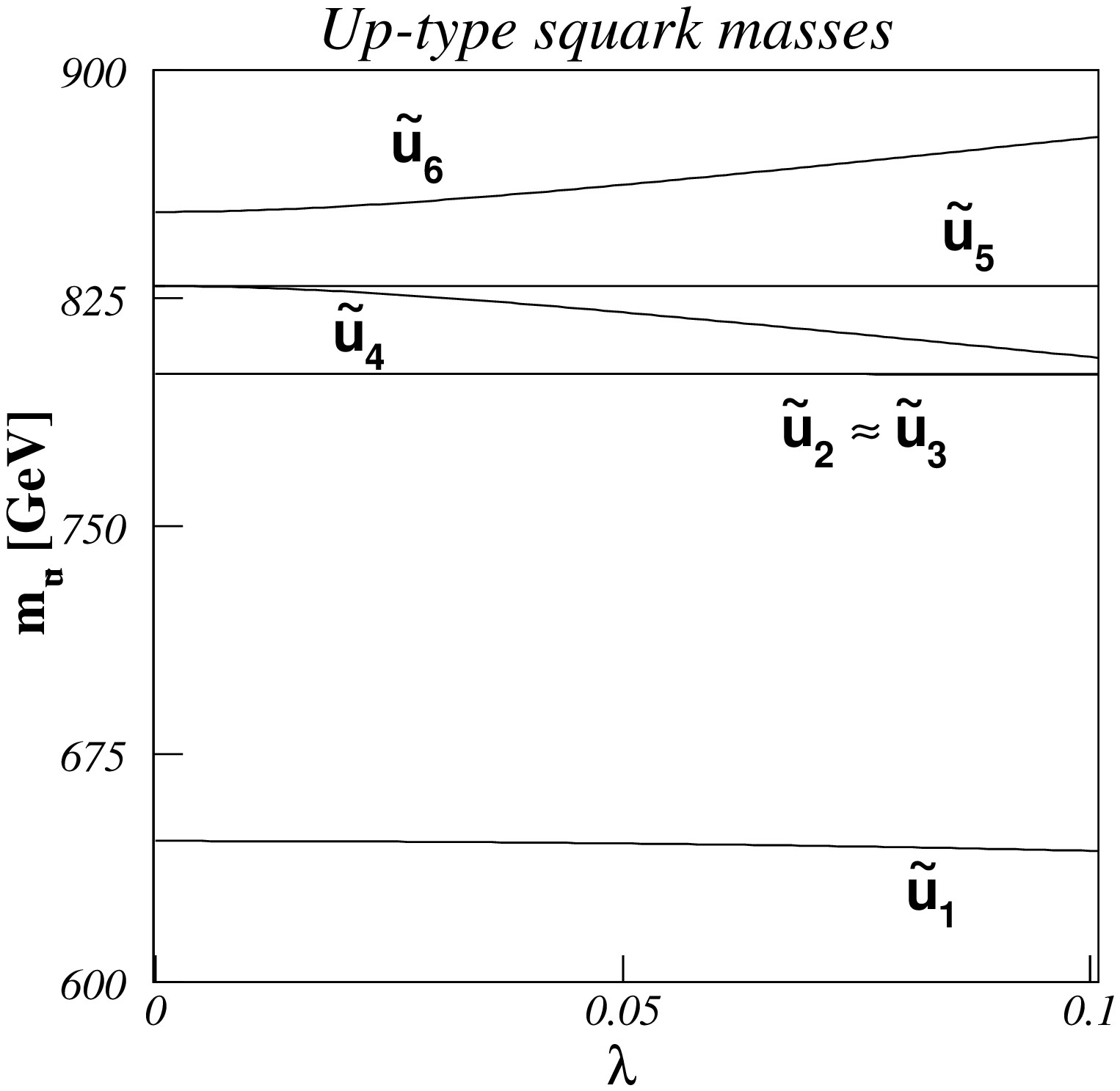}
    \hfill
    \includegraphics[scale=0.23]{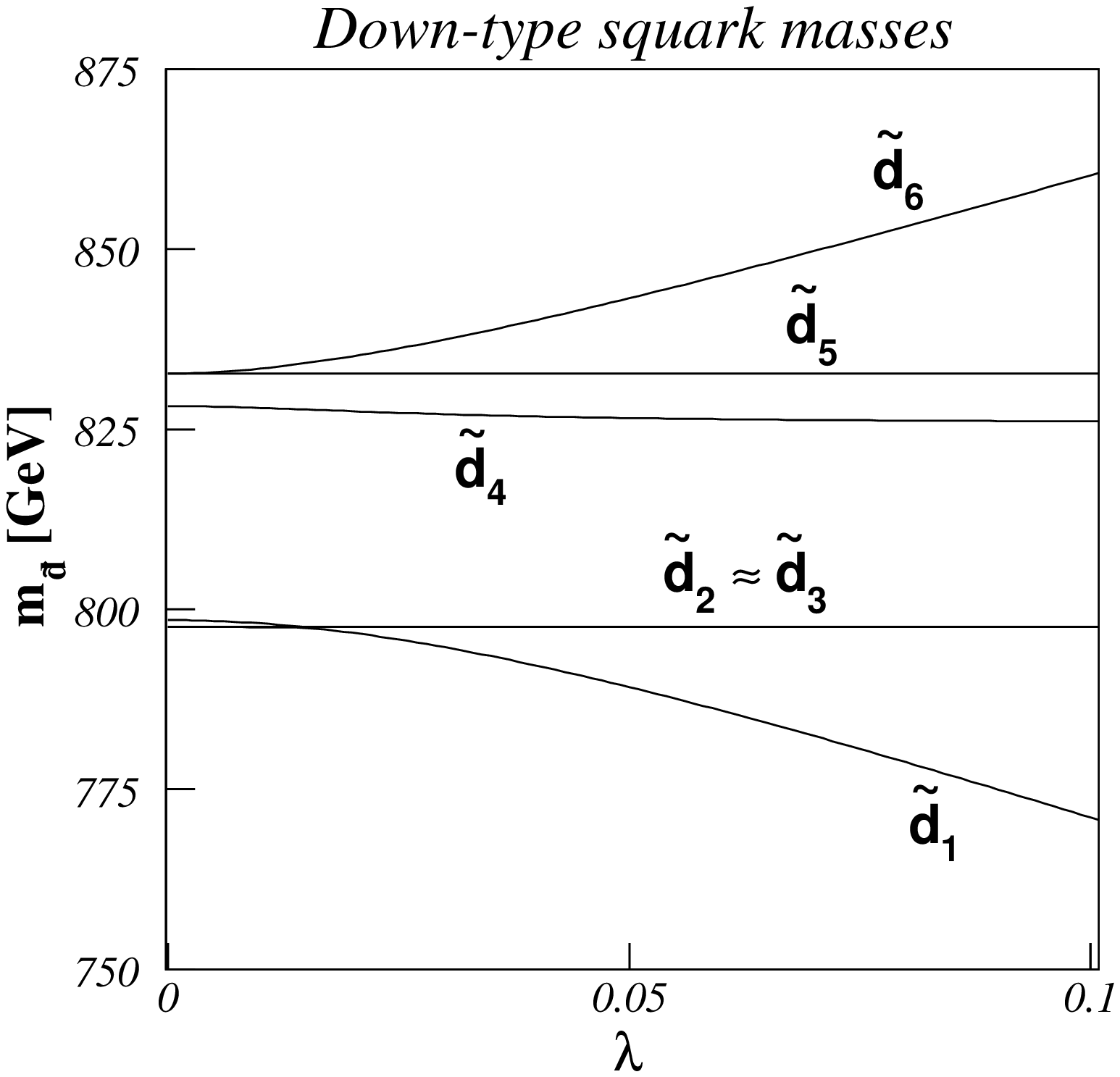}
\caption{Dependence of the up- and down-type squark masses on the NMFV-parameter
$\lambda$ in our benchmark scenario B.} 
\label{fig3}
\end{figure}

The explicitly imposed constraints are shown in Fig. \ref{fig2}, except for
the SUSY contribution to the anomalous magnetic moment of the muon, 
$a_{\mu}^{\rm SUSY} \simeq 14\times 10^{-10}$, that is independent of $\lambda$
and agrees with the limits in Eq. (\ref{eqamu}). The first graph 
shows the electroweak precision observable $\Delta\rho$. On our logarithmic
scale, only the upper bound of the 2$\sigma$ range is visible. Despite the
strong dependence of $\Delta\rho$ on squark flavours, helicities and masses, the
experimental error is sufficiently important to allow for relatively large values of
$\lambda \le 0.52$. The most stringent low-energy constraint comes from the
$b\to s\gamma$ branching ratio, which depends strongly on the NMFV-parameter
$\lambda$ due to the same loop-level of the SUSY and SM contributions. The small
experimental error allows for two narrow intervals in $\lambda$. As the
second one is disfavoured by $B\to X_s\mu^+\mu^-$ \cite{ref52}, we restrict
ourselves to the region $\lambda \le 0.1$. The neutralino relic density
$\Omega_{\rm CDM}h^2$ agrees with the limits in Eq. (\ref{eqcdm}) up to high
values of $\lambda \le 0.9$, which is due to the fact that in the relevant
neutralino (co)annihilation processes squarks only appear as propagators or at
one-loop level. The decreasing of $\Omega_{\rm CDM}h^2$ for higher $\lambda$ is due
to the larger mass splitting of the squarks, implying that the lightest up- and
down-type squarks become lighter and coannihilation processes become more
important.

\begin{figure*}
    \includegraphics[scale=0.226]{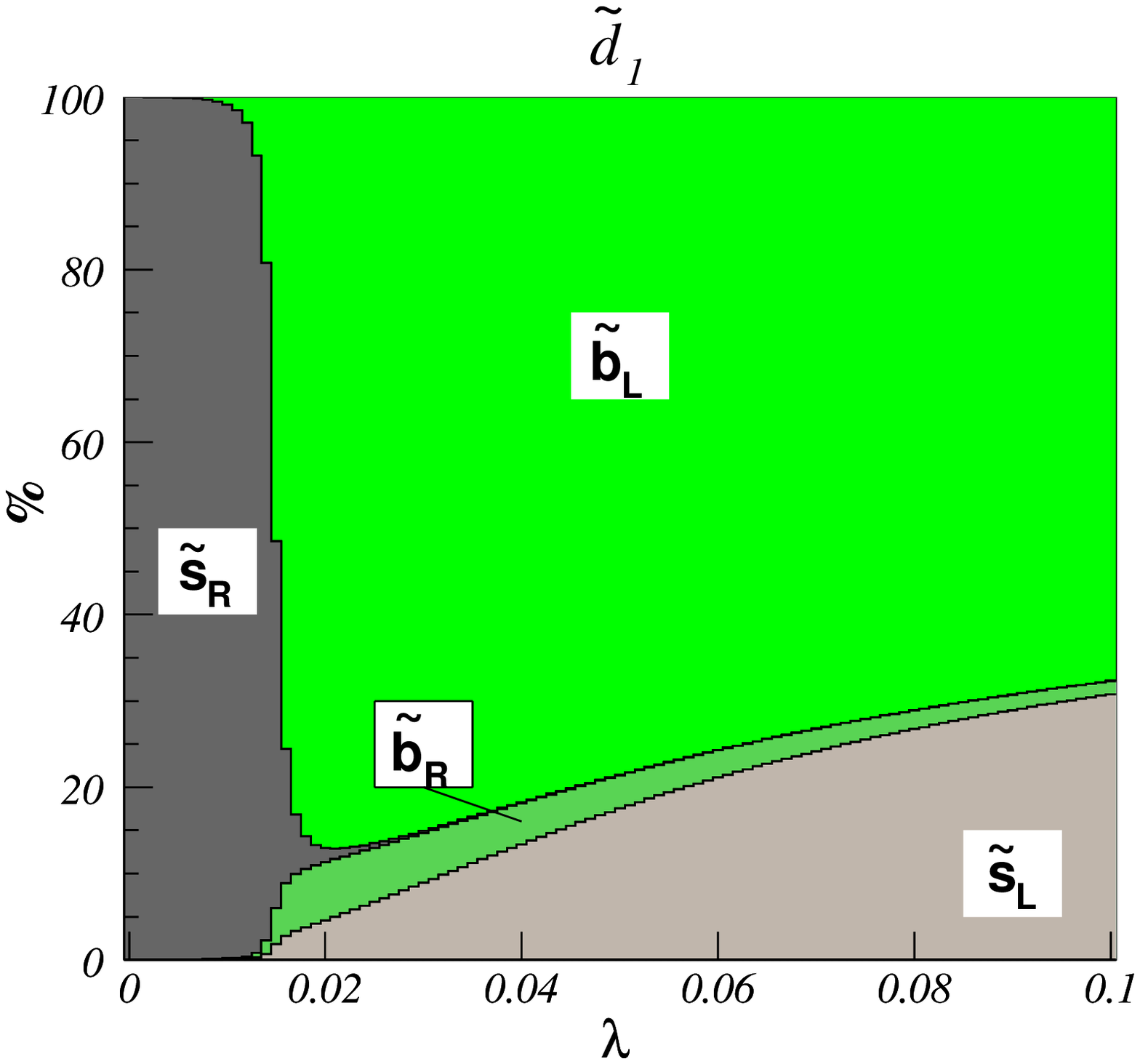}
    \hfill
    \includegraphics[scale=0.226]{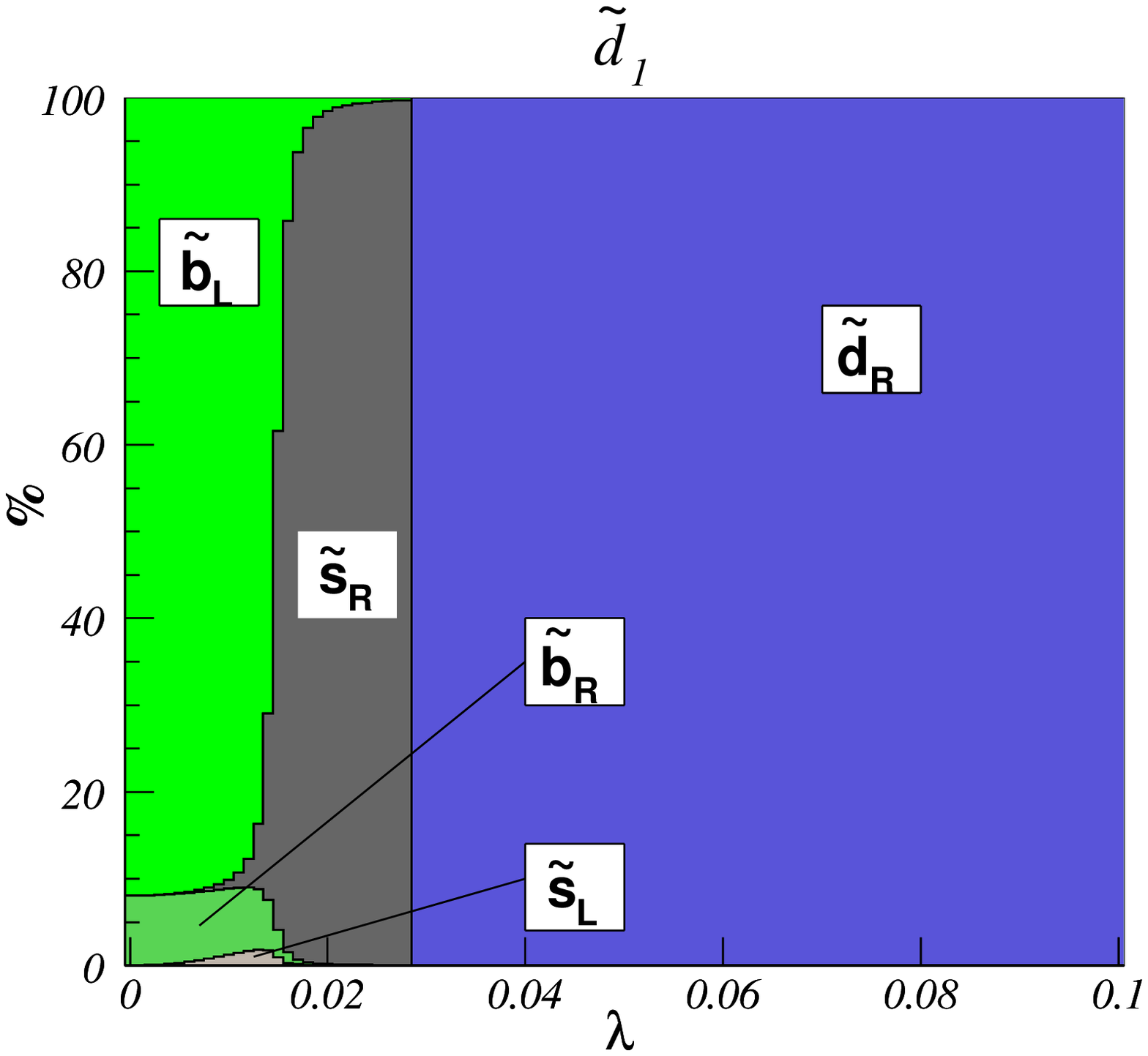}
    \hfill
    \includegraphics[scale=0.226]{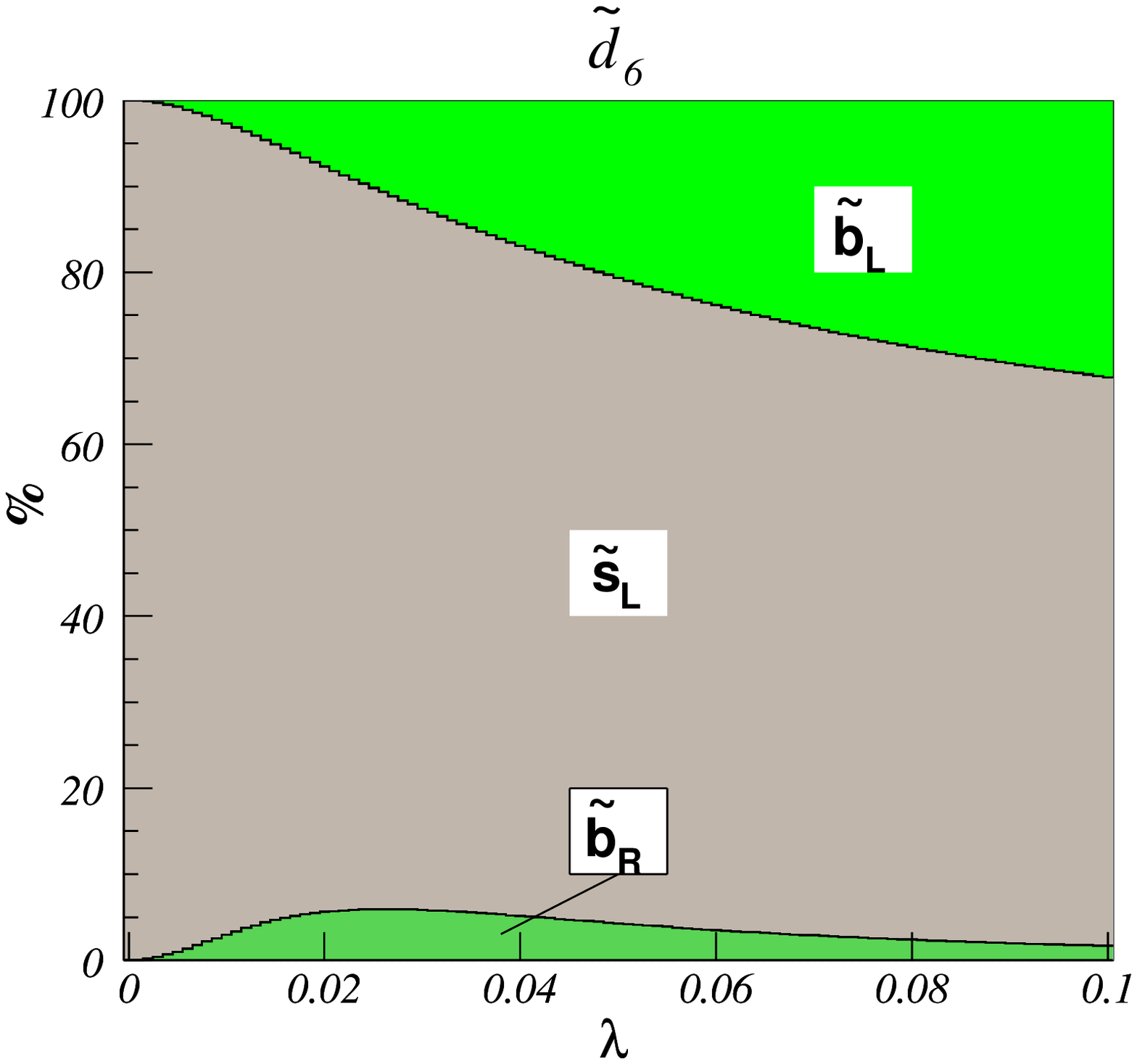}
    \hfill
    \includegraphics[scale=0.226]{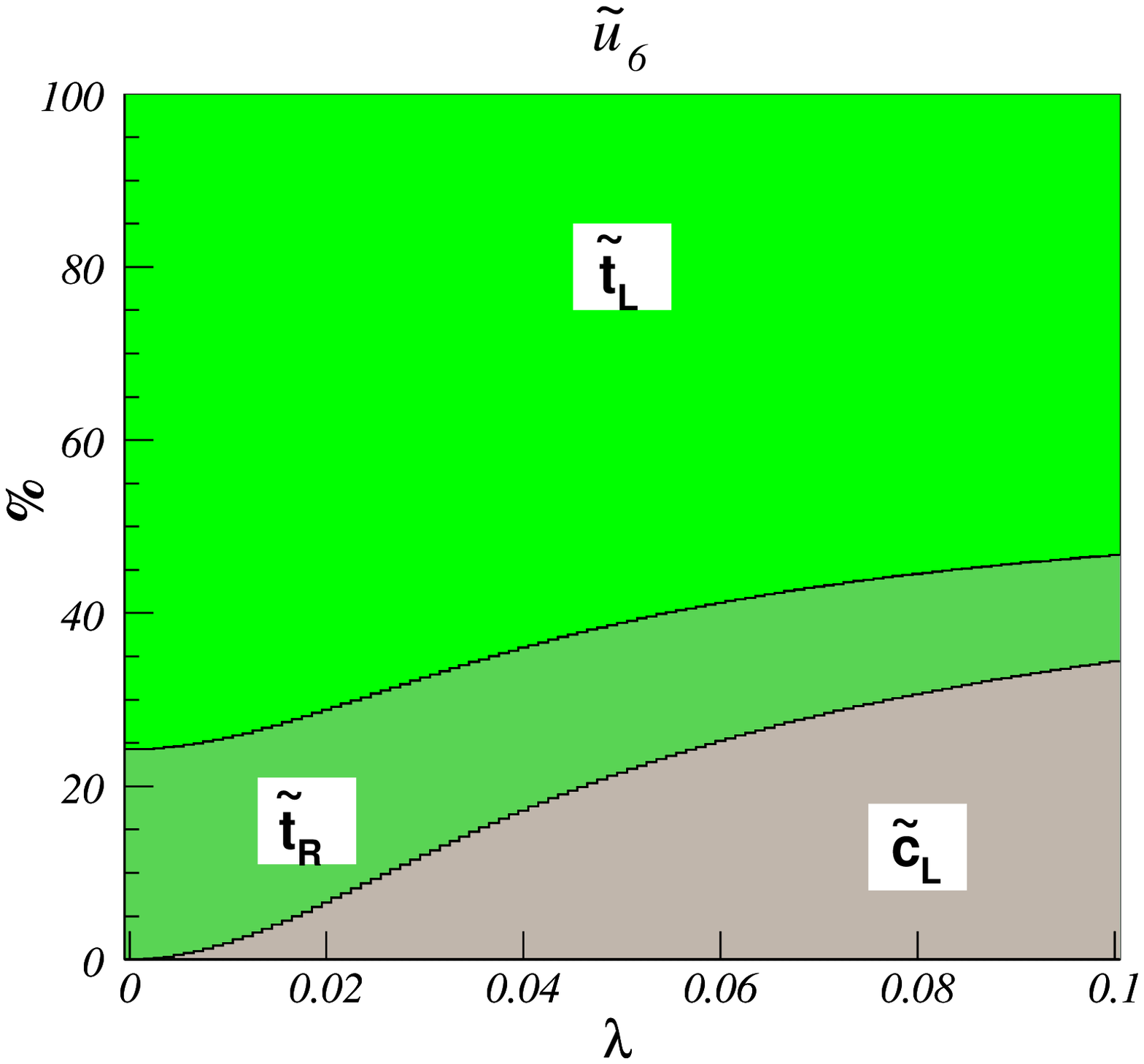}
\caption{Dependence of the chirality (L, R) and flavour ($u$, $c$, $t$; $d$,
$s$, and $b$) content of the discussed up- and down-type squark mass eigenstates
on the NMFV-parameter $\lambda\in[0;0.1]$ for benchmark scenario B.}  
\label{fig4}
\vspace*{5mm}
\end{figure*}

Fig. \ref{fig3} shows the mass eigenvalues of the up- and down-type squarks
as a function of the NMFV parameter $\lambda \in [0;0.1]$. We observe large mass
splitting for the lightest and heaviest squarks with increasing $\lambda$ due
the growing relative importance of the off-diagonal matrix elements. However,
the masses of the intermediate squarks are practically insensitive to $\lambda$.
Another interesting phenomenon observed is the so-called ``avoided crossing"
between the squark mass eigenvalues. For the down-type squarks for example, this is
observed at $\lambda \approx 0.02$ between the states $\tilde{d}_1$ and
$\tilde{d}_3$ (see Fig. \ref{fig3}). At the point where two levels should cross,
the corresponding squark eigenstates mix and change character (see Fig. \ref{fig4}). 
The same phenomenon occurs also for the up-type squarks at $\lambda \approx 0.02$
between $\tilde{u}_4$ and $\tilde{u}_5$, and a second time at $\lambda \approx
0.1$ between $\tilde{u}_3$ and $\tilde{u}_4$. These ``avoided crossings", a 
common phenomenon in quantum mechanics, are due to the fact that the Hermitian 
squark mass matrix depends continuously on one single real parameter $\lambda$.

In Fig. \ref{fig4} we show the helicity and flavour decomposition of some of the
up- and down-type left-handed squark mass eigenstates for the experimentally 
fa\-vou\-red range in the vicinity of (c)MFV, i.e. $\lambda \in [0;0.1]$. 

\begin{figure*}
    ~\includegraphics[scale=0.22]{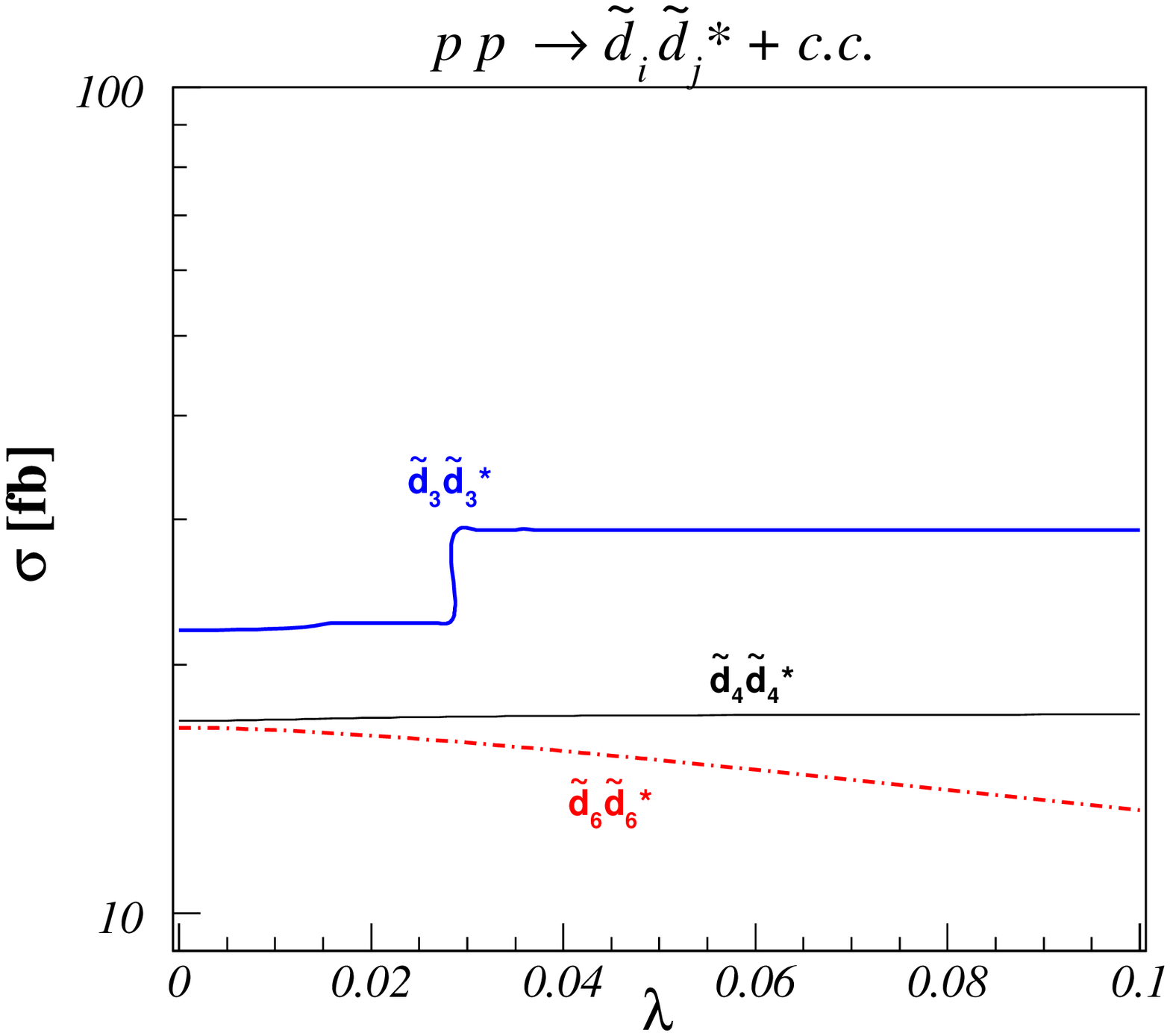}
    \hfill
    \includegraphics[scale=0.22]{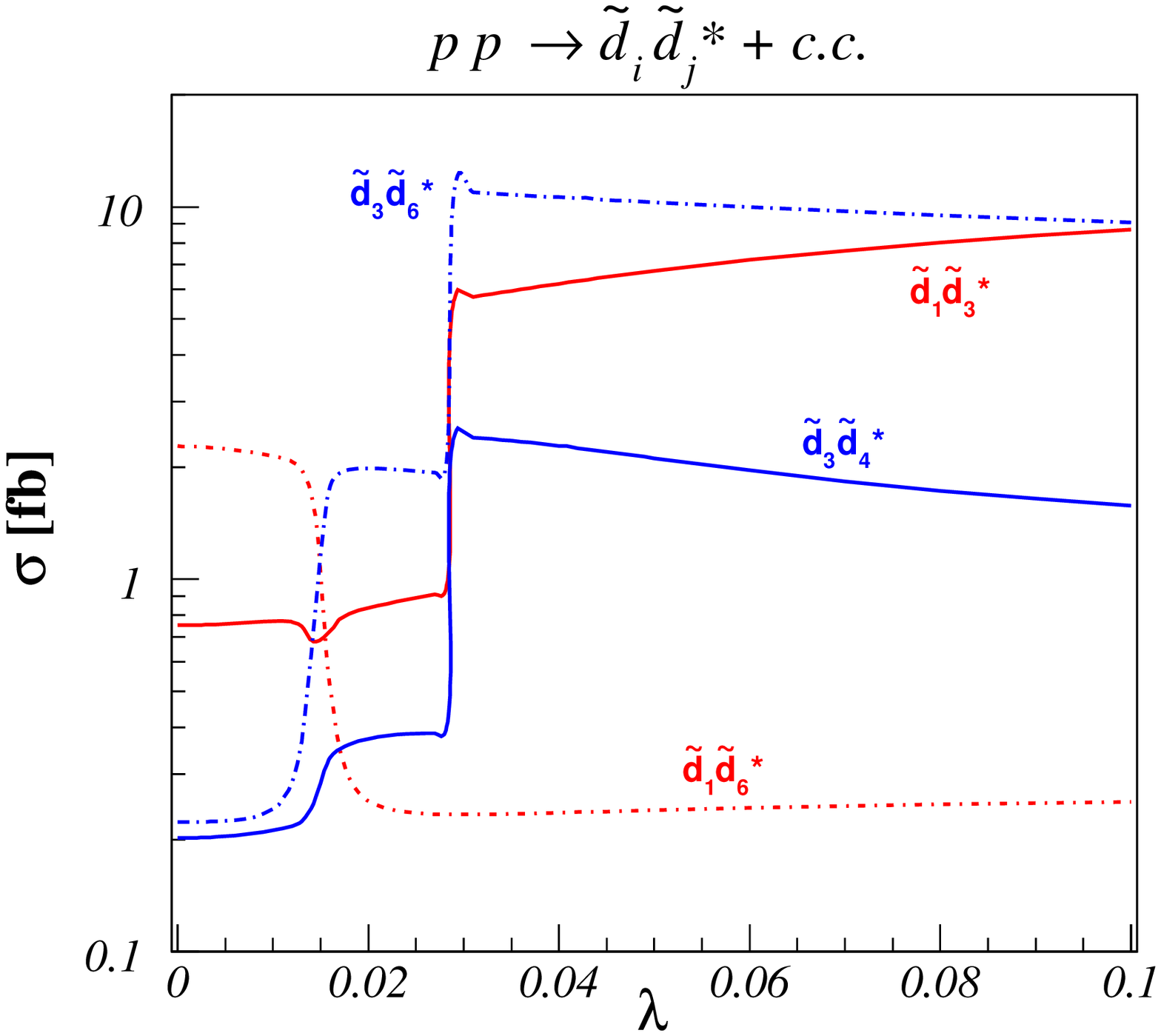}
    \hfill
    \includegraphics[scale=0.22]{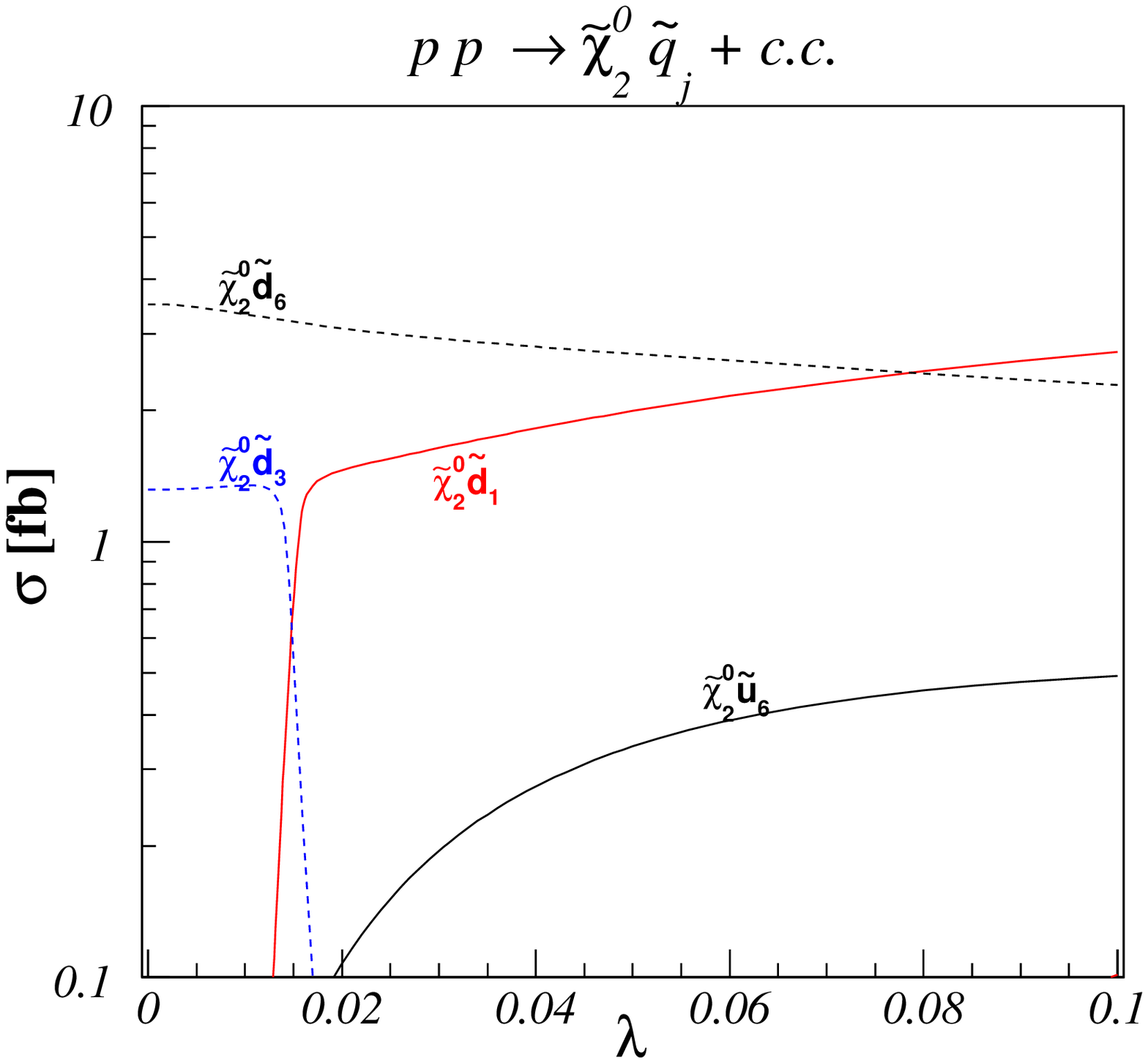}
    \hfill
    \includegraphics[scale=0.22]{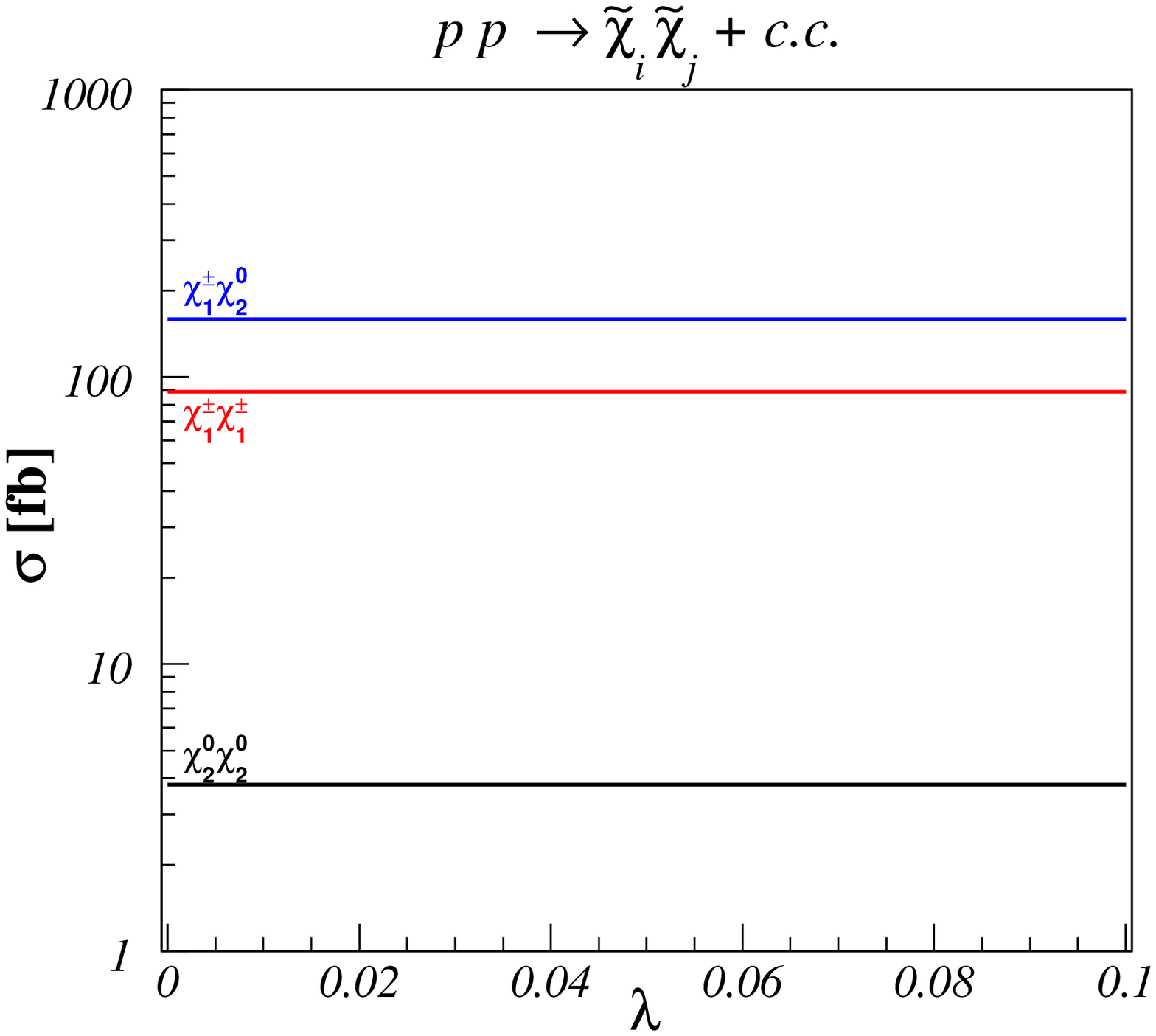}
\caption{Cross section examples for charged and neutral squark-antisquark pair,
associated squark-neutralino, and gaugino-pair production at the LHC as function
of the NMFV-parameter $\lambda \in [0;0.1]$ in our benchmark scenario B.}  
\label{fig5}
\end{figure*}

We finally provide numerical predictions for production cross sections of
squarks and gauginos at the LHC, i.e. for $pp$-collisions at $\sqrt{s}=14$ TeV
centre-of-momentum energy. Fig. \ref{fig5} shows the numerical results for
some interesting channels in our benchmark scenario B. Again, for a complete
discussion of all channels in all four benchmark scenarios, the reader is
referred to Ref. \cite{ref0}. 
The magnitudes of the cross sections vary from barely visible level ($10^{-2}$
fb) for weak production of heavy final states over the semi-strong production of
squarks and gauginos and quark-gluon final states to large cross
sections ($10^2$ to $10^3$ fb) for the strong production of diagonal
squark-(anti)squark pairs or weak production of very light gaugino pairs.
Unfortunately, since the strong gauge interaction is insensitive to (s)quark
flavours and gaugino pair production cross sections are summed over exchanged
squark  flavours, the processes with the largest cross sections are practically
insensitive to the flavour violation parameter $\lambda$.

Some of the subleading, non-diagonal cross sections, however, show sharp
transitions. For example, at $\lambda=0.02$, the cross sections for
$\tilde{d}_1\tilde{d}_6^*$ and $\tilde{d}_3\tilde{d}_6^*$ switch places,
corresponding to an avoided crossing in the mass eigenvalues and a switch in the
flavour contents of the $\tilde{d}_1$ and $\tilde{d}_3$ squarks (see Figs.
\ref{fig3} and \ref{fig4}). Since the concerned sstrange and sbottom mass
differences are rather small, this is mainly due to the different strange and
bottom quark densities in the proton. At $\lambda=0.035$, the cross sections for
$\tilde{d}_3\tilde{d}_6^*$ and $\tilde{d}_1\tilde{d}_3^*$ increase sharply,
corresponding to the flavour content change in the $\tilde{d}_3$ squark from
$\tilde{s}_R$ to $\tilde{d}_R$ (see Fig. \ref{fig4}), that can now be produced
from down valence quarks. The $\tilde{d}_1\tilde{d}_3$ production cross section
increases with the strange squark content of $\tilde{d}_1$.

Smooth transitions and semi-strong cross sections of about 1 fb are observed for
the associated production of third-generation squarks with charginos and
neutralinos. The cross section for $\tilde{d}_6$ production decreases with its
strange squark content, while the bottom squark content increases at the same
time (see Fig. \ref{fig4}). The opposite happens for $\tilde{d}_3$, where the
production cross section increases. For the up-type squarks, the cross section
of $\tilde{u}_6$ production increases/decreases with its charm/top squark
content.

\section{Conclusion}

We have presented an implementation of NMFV in the MSSM at the weak scale. The
relevant flavour violating off-diagonal elements of the squared squark mass
matrix are parametrized by one real NMFV parameter. Based on an extensive
analysis of low-energy, electroweak precision and cosmological constraints, we
proposed benchmark points for the mSUGRA scenario including NMFV. We discussed
the phenomenology of these benchmark scenarios, which are characterized by the
phenomenon of ``avoided crossings". We finally presented numerical predictions
for squark and gaugino production cross sections at the LHC, among which several
processes are sensitive to NMFV.

\end{document}